\documentclass[]{aastex}
\usepackage{emulateapj5,ifthen}
\shorttitle{Effects of a Major Merger on Abell 2319}
\shortauthors{O'Hara, Mohr, \& Guerrero}

\newcommand{\kev}{{\rm\ keV}}
\newcommand{\Chandra}{{\it Chandra}}
\newcommand{\ROSAT}{{\it ROSAT}}
\newcommand{\ASCA}{{\it ASCA}}
\def\Micm{$M_{\rm ICM}$}
\def\L2500{$L_{2500}$}
\def\M2500{$M_{2500}$}
\def\r2500{$r_{2500}$}
\def\rfive{$r_{500}$}
\def\RI{$R_{\rm I}$}
\def\Tx{$T_{\rm X}$}
\def\NH{$N_{\rm H}$}

\newcommand{\figtype}{EPS}
\def\myputfigure#1#2#3#4#5%
{\vskip#5pt\makebox[0pt]{\hskip#2in
\includegraphics[width=#3\textwidth]{#1}}\vskip#4pt\hfill}
\newenvironment{inlinefigure}{
\def\@captype{figure}
\noindent\begin{minipage}{0.999\linewidth}\begin{center}}
{\end{center}\end{minipage}\smallskip}

\begin{document}

\submitted{Submitted to ApJ August 22, 2003}

\title{A {\it Chandra} Study of the Effects of 
a Major Merger\\ on the Structure of Abell 2319}

\author{Timothy B. O'Hara\altaffilmark{1,2}, Joseph J. Mohr\altaffilmark{2,1},
  and Mart\'{\i}n A. Guerrero\altaffilmark{2,3}}

\altaffiltext{1}{Department of Physics, University of Illinois,
1110 West Green St, Urbana, IL 61801; tbohara@astro.uiuc.edu, jmohr@uiuc.edu}
\altaffiltext{2}{Department of Astronomy, University of Illinois,
1002 West Green St, Urbana, IL 61801}
\altaffiltext{3}{Current Address: Instituto de Astrof\'{\i}sica de
Andaluc\'{\i}a, CSIC, Spain; mar@iaa.es}

\begin{abstract}

We present an analysis of a \Chandra\ observation of the massive, nearby galaxy
cluster Abell 2319.   A sharp surface brightness discontinuity---suggested by
previous, lower angular resolution X-ray imaging---is clearly visible in the ACIS image.  This $\sim$300~kpc  feature suggests
that a major merger is taking place with a significant velocity component 
perpendicular to the line of sight.  
The cluster emission-weighted
mean temperature is $11.8 \pm 0.6 \kev$, somewhat higher than previous
temperature measurements.  The \Chandra\ temperature map of A2319  reveals substructure 
resembling that anticipated based on hydrodynamic simulations of cluster 
mergers, and shows an associated cool core not previously known. The map
shows a separation between the intracluster medium (ICM) and galaxies of one subcluster,
indicating a transient state in which the ICM has been stripped from the
subcluster galaxies (and presumably the dark matter).
Detailed analysis of the merger feature shows a pressure change across the surface brightness discontinuity by a factor of $\lesssim 2.5$.  The higher density side of the front has a lower
temperature, suggesting the
presence of a cold front similar to those in many other merging clusters. 
The velocity of the front is roughly sonic.

We compare bulk properties of the ICM and galaxies in A2319 to the same properties in a large sample of clusters as a way of gauging the effects
of the major merger. Interestingly, by comparing A2319 to a sample of 44 clusters studied with the \ROSAT\ PSPC we find that the X-ray luminosity, isophotal size, and ICM mass  are consistent with the expected values for a cluster of its temperature; 
in addition,
the $K$-band galaxy light is consistent with the light--temperature scaling relation derived from a sample of $\sim$100 clusters studied with 2MASS.  
Together, these results
indicate either that the merger in A2319 has not been effective at altering the bulk properties of the cluster, or that there are large but correlated displacements in luminosity, isophotal size, ICM mass, galaxy light, and emission-weighted mean temperature in this cluster.

\end{abstract}

\keywords{galaxies: clusters: general ---
galaxies: clusters: individual (Abell 2319) --- 
X-rays: galaxies: clusters
}

\section{Introduction}
\label{sec:intro}

Galaxy cluster mergers are highly energetic events, driving shocks into
the intracluster medium (ICM) of the colliding clusters.   Flattened and asymmetric X-ray morphologies are signatures of recent merging \citep{mohr93}, and these signatures have been used to study the prevalence of merging in large samples of present-epoch clusters \citep{mohr95,buote96}. A study of X-ray images of a flux-limited sample of 65 clusters indicates that more than half of nearby clusters show evidence of merging \citep{mohr95}. Hydrodynamical simulations
indicate that complex temperature structures should also be produced in these
mergers; however, until relatively recently the required spectral and angular resolution to map this structure has not been available. 
\Chandra\ and {\it XMM-Newton}, with their high angular resolution, are well-suited for
detailed studies of merger features in galaxy clusters
\citep[e.g., ][]{markevitch00,vikhlinin01,markevitch01b,
sun02,markevitch02,kempner02,maughan03}. These studies
have already revealed that merger features observed
in clusters may not indicate shock fronts, but rather ``cold fronts,'' 
wherein the cool, dense cores of clusters survive through the initial
impact of the merger \citep{markevitch00}. In fact, it now appears
that many well-known merger candidates contain these cold fronts, e.g., 
A2142 \citep{markevitch00}, A3667 \citep{vikhlinin01}, 
A2256 \citep{sun02}, and A85 \citep{kempner02}.

Abell 2319 is a massive nearby cluster \citep[$z=0.0564$;][]{abell58,struble87}.  We chose to study it with the high resolution of \Chandra\ because its X-ray morphology observed at lower resolution with the \ROSAT\ PSPC shows a strong asymmetry or ``centroid variation,'' which is a classic indicator of a recent merger.  Our goal in this study is not only to better understand the merger state of A2319, but also to determine how the ongoing merger in A2319 is affecting its bulk ICM and galaxy properties.  
Of particular interest is understanding how merging---which has long been known to be prevalent in the cluster population \citep{geller82,dressler88,mohr95}---is likely to impact attempts to use cluster surveys to study cosmology \citep[e.g., ][]{haiman01,randall02,majumdar03a,majumdar03b,hu03b}.

On the basis of galaxy spectra, \citet{faber77} suggested that A2319 is
actually composed of two clusters superimposed along the line of sight, 
with the smaller subcluster located $\sim 10\arcmin$ to the northwest of the
main cluster and X-ray surface brightness peak. Additional redshift 
measurements led to an estimated mean velocity for the main subcluster
of $\sim 100$ members (hereafter A2319A) of 15727 km s$^{-1}$, and for
the smaller subcluster of $\sim 25$ members (hereafter A2319B) of 
18636 km s$^{-1}$ \citep{oegerle95}. This analysis suggests that there is a $\sim 50\%$ chance 
that the two subclusters are in fact not gravitationally bound and will not merge. 

A2319 has been extensively studied with previous X-ray instruments, and the inferences about the cluster dynamical state have been quite varied. Emission-weighted mean temperature estimates are generally in the 9--10 keV range \citep[e.g., ][]{david93, markevitch98, molendi99, irwin00}.   \citet{markevitch96} produced a temperature map of A2319 using \ASCA.
These observations provided no evidence for a cold core region near
the surface brightness discontinuity, though a region to
the northwest of the brightness peak appeared to have a temperature  $\sim 1.5 \kev$ lower than the mean.  This same subcluster region was identified by
\citet{molendi99} using {\it BeppoSAX}; it is proposed that this cool region is
associated with subcluster A2319B.   
Using the \ASCA\ temperature map, \citet{markevitch96}
argued that there is no evidence of a large-scale merger in A2319. 
\citet{mohr95}, however, found a value for the centroid variation of A2319 in the {\it Einstein} IPC image
that indicates an ongoing merger.  Interestingly, a combined X-ray and
radio study of the cluster suggests
that the two subclusters are in a premerger state \citep{feretti97}.  This study also takes note of X-ray evidence for another merger in a late stage taking place
along the northeast-southwest direction.

\vskip20pt
\begin{inlinefigure}
   \ifthenelse{\equal{\figtype}{EPS}}{
   \begin{center}
   \epsfxsize=8.cm
   \begin{minipage}{\epsfxsize}\epsffile{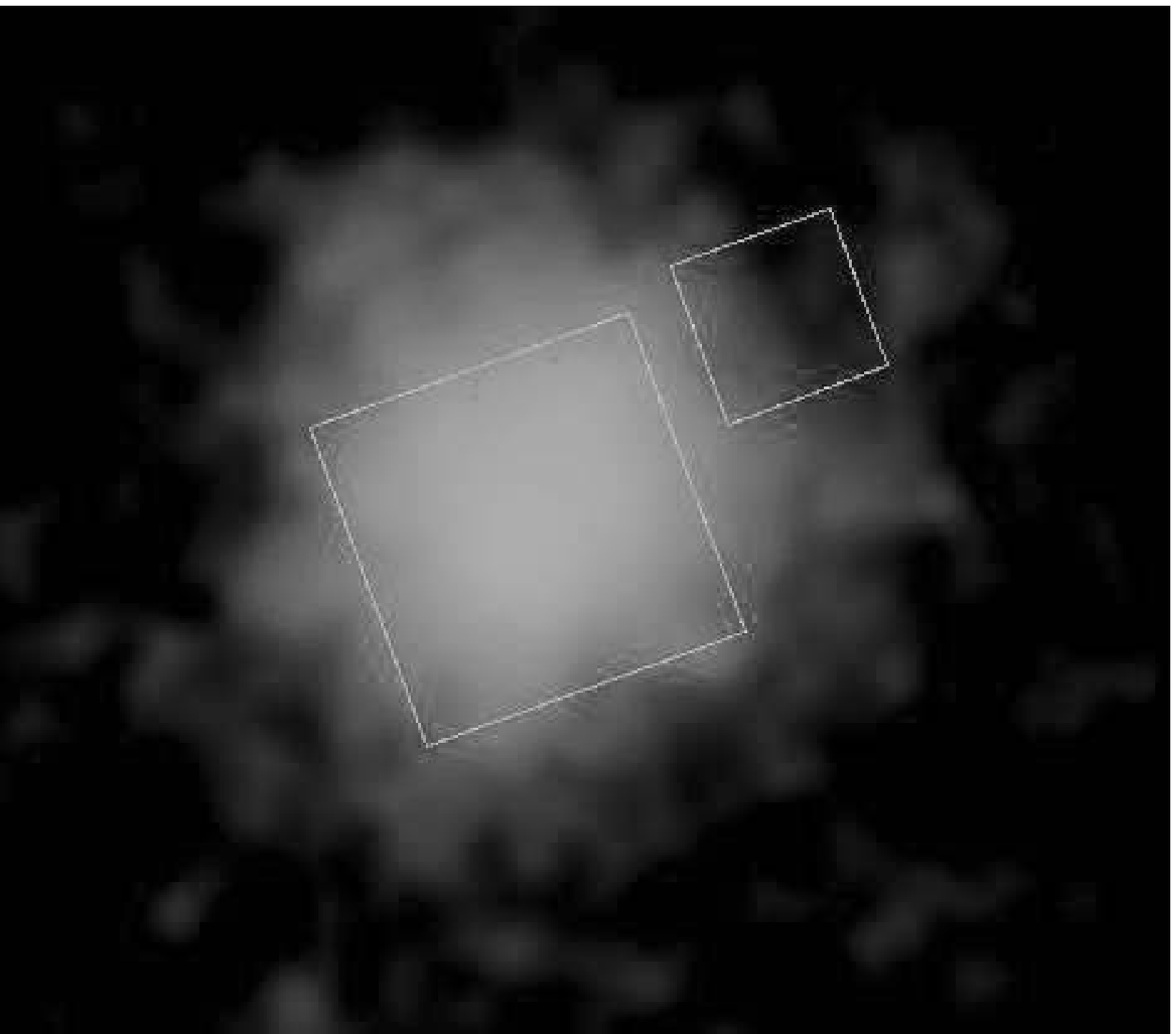}\end{minipage}
   \end{center}}
   {\myputfigure{rosat.pdf}{1.0}{1.1}{-0}{-180}}
   \figcaption{\label{fig:rosat} 
   	\ROSAT\ PSPC image of A2319 with \Chandra\ observation footprint
	overlaid. North is up and east is to the left in all images.
	The ACIS-I footprint is roughly 17$\arcmin$ on a side.
     }
\end{inlinefigure}

In this paper, we present a detailed X-ray study of A2319 based on imaging spectroscopy
from \Chandra\ ACIS-I, providing clear evidence for an ongoing
merger of  two major subclusters. In \S\ref{sec:obs} we present the observations. After
a description of the data reduction process in \S\ref{sec:reduction}, we
present an analysis of the overall cluster spectrum (\S\ref{sec:spec}) and a 
temperature map of the cluster (\S\ref{sec:temp}). In \S\ref{sec:merger} we
analyze the merger feature in detail, including quantitative estimates
of changes in the physical state of the ICM across the feature, and propose
a simple dynamical model.
This is followed in
\S\ref{sec:structure} by a study of how this merger has affected the bulk X--ray properties of the cluster;  we examine how A2319---a cluster in the middle of a major merger---behaves relative to an X-ray flux-limited sample of clusters in its luminosity, isophotal size, and ICM mass.
Finally, we summarize our conclusions in \S\ref{sec:concl}.

Throughout the paper we assume a $\Lambda$CDM cosmology with $\Omega_M=0.3$ and 
$\Omega_\Lambda = 0.7$, and take the Hubble parameter to be
$H_0 = 70\,h_{70}$~km~s$^{-1}$~Mpc$^{-1}$.

\section{Observation}
\label{sec:obs}


A2319 was observed with \Chandra\ on 15 March 2002 for 14.6 ks using
ACIS-I and ACIS-S2, with the ACIS-I field of view centered at 
$\alpha$ = 19h21m12.00s, $\delta$ = +43$^{\circ}$56$\arcmin43.7\arcsec$,
roughly on the surface brightness peak. 
The pixel scale is $0\arcsec.492$.
Time bins were checked for periods
with count rates greater or less than 20\% of the mean; no such intervals
were found. Hence all of the data with grades of 0, 2, 3, 4, and 6 were
used. The ACIS-I data were adjusted for charge-transfer inefficiency (CTI)
using the PSU CTI corrector \citep{townsley00}.
We used the \Chandra\ data analysis software CIAO, version 2.2, for 
data reduction. All spectral analysis was done using
the X-ray spectral fitting package XSPEC, version 11.2.

\vskip20pt
\begin{inlinefigure}
   \ifthenelse{\equal{\figtype}{EPS}}{
   \begin{center}
   \epsfxsize=8.cm
   \begin{minipage}{\epsfxsize}\epsffile{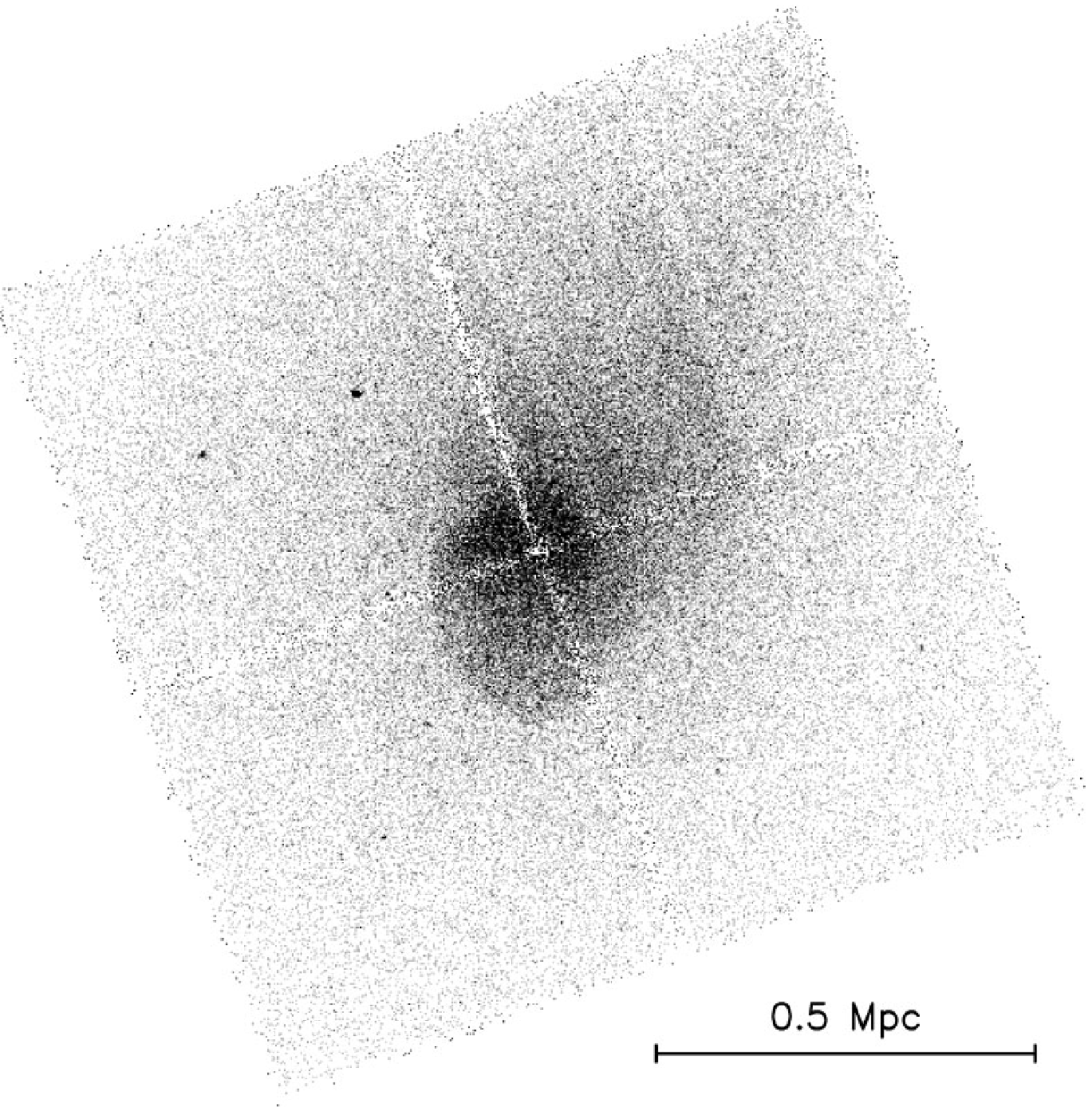}\end{minipage}
   \end{center}}
   {\myputfigure{counts.pdf}{1.7}{1.5}{-10}{-250}}
   \figcaption{\label{fig:counts} 
   Raw counts ACIS-I image in the 0.5--5 keV band, 
	with pixels binned by 4 (i.e., 
	the pixel scale is $\sim 2\arcsec$). The merger feature is visible to 
	the southeast of the brightness peak, and the ``tail'' of diffuse
	emission is seen extending to the northwest.
     }
\end{inlinefigure}

\subsection{Background and Imaging}
\label{sec:reduction}

Because A2319 is a large, nearby cluster, its emission fills the ACIS-I chip,
preventing a direct  background measurement from that data set. The count rate
in the S2 chip is found to be roughly two times higher than the typical
background rate, making its use for background estimation likewise dubious. 
One source of this higher than expected rate could be a flare affecting
the entire observation; however, the uniformity of the count rate over the
 exposure time makes this unlikely, and a visual inspection of 
the S2 spectrum does not reveal any flare-like features. A clear 
brightness gradient is visible in the exposure-weighted S2 image,
as well as in the \ROSAT\ PSPC image shown with the \Chandra\ footprint
in Figure~\ref{fig:rosat};
hence, it is clear that emission from the very extended cluster
is present in the S2 data. 

Because there were no portions of the observation without significant cluster contamination, we use the Markevitch blank-sky observations\footnote{http://cxc.harvard.edu/contrib/maxim/acisbg/}.
The background was scaled up by $\sim$10\% after visual comparison of the S2 spectrum and the blank sky spectrum, under the assumption that emission in the 7--10 keV band is background dominated. The recommended procedure for using these blank-sky files is to compare the emission
in the 10--12 keV band; however, the spectral shapes of the S2 spectrum and the background
spectrum are somewhat better matched in the 7.0--10 keV band, and matching the two spectra
in the higher band results in obvious oversubtraction at energies below 10 keV.
We compared the blank-sky corrected mean surface brightness in the
S2 data to that of the background corrected PSPC observation; the {\it Chandra} measured surface brightness is brighter by a factor of $\sim1.5$.

The raw ACIS-I exposure-weighted counts image in the 0.5-5.0 keV band
 is shown in Figure~\ref{fig:counts}. The presumed merger feature is visible
as a sharp surface brightness discontinuity to the southeast of the
brightness peak. The presence of the merger signature is much clearer than
in previous X-ray observations; the arclike discontinuity and the ``tail''
of emission towards the northwest strongly resemble similar features in
merging clusters such as A2142 and A3667. This is not the possible merger in the
northeast-southwest direction discussed by, e.g., \citet{feretti97}, as it
clearly indicates gas movements along the axis connecting A2319A and A2319B.

\subsection{Spectral Analysis}
\label{sec:spec}

All spectra are fitted using a single-temperature MEKAL model, plus
components for absorption along the line-of-sight and for absorption due to
molecular contamination of
the ACIS detector. We fit spectra in the
energy range 0.9--10.0 keV; poor understanding of the low-energy response
of ACIS-I prevents us from using data at lower energies.

We first fit for temperature
and abundance, fixing the hydrogen column density
at the \citet{dickey90} value of $8.33\times10^{20}$ cm$^{-2}$.
Fitting over the entire cluster, excluding point sources, 
gives $T_{\rm X} = 11.8 \pm 0.2 \kev$ and 
$Z = 0.19 \pm 0.03$ (all abundances are in units of solar abundance; 
all fitted uncertainties are at the 1 $\sigma$ level),
with $\chi^2 = 1017$ for 594 degrees of freedom.
This temperature is several standard deviations above previously published
estimates, e.g., $T_{\rm X} = 9.2 \pm 0.7 \kev$ determined by
 \citet{markevitch98} using {\it ASCA} data.
This spectrum is plotted
with residuals in Figure~\ref{fig:allspectrum}.

Previous studies of A2319 have used hydrogen column densities in the 
range $(7.85-8.9)\times10^{20}$ cm$^{-2}$; often the value of \NH\ used
is not provided. 
By fitting the entire cluster spectrum with varying values
of \NH, we have found that the emission-weighted mean temperature
varies roughly linearly with \NH, with the temperature decreasing
by approximately 0.5 keV per 10$^{20}$ cm$^{-2}$ (cluster temperature uncertainties
are generally $\sim$0.2 keV).
Fitting for the column density along with the other
parameters yields $T_{\rm X} = 10.6 \pm 0.3$ keV, $Z = 0.20 \pm 0.03$, and
$N_{\rm H} = (10.7 \pm 0.5)\times10^{20}$ cm$^{-2}$,
with $\chi^2 = 999$ for 593 degrees of freedom.

The \citet{dickey90} value for the hydrogen column density
of $8.33\times10^{20}$ cm$^{-2}$, as well as other values
used in previous studies of A2319, fall a few standard deviations
below the range of our fit value.
However, uncertainties are not readily available for the H{\sc i} 
survey data of
\citet{dickey90}; moreover, measured \NH\ values in the region of the
sky around A2319 vary to levels above our fit value. A2319 lies at a fairly
low galactic latitude where there is a significant amount of ISM along the line
of sight, and the optically thin assumption for deriving \NH\ likely underestimates
the true column density by a factor of 1.1--1.3 \citep{dickey90}. Further, with a 
column density this high there is likely to be a significant contribution ($\geq10\%$) to
the hydrogen column by molecular hydrogen \citep{lockman03}.
Also, fitting \NH\ along
with other parameters in our temperature mapping 
suggests that there may be a gradient with magnitude of a few 10$^{20}$ 
cm$^{-2}$ across the ACIS-I image. 
For the rest of the paper we adopt the value of $8.33\times10^{20}$ cm$^{-2}$, but readers should keep in mind that it is almost certainly too low. 

\vskip20pt
\begin{inlinefigure}
   \ifthenelse{\equal{\figtype}{EPS}}{
   \begin{center}
   \epsfxsize=8.cm
   \begin{minipage}{\epsfxsize}\epsffile{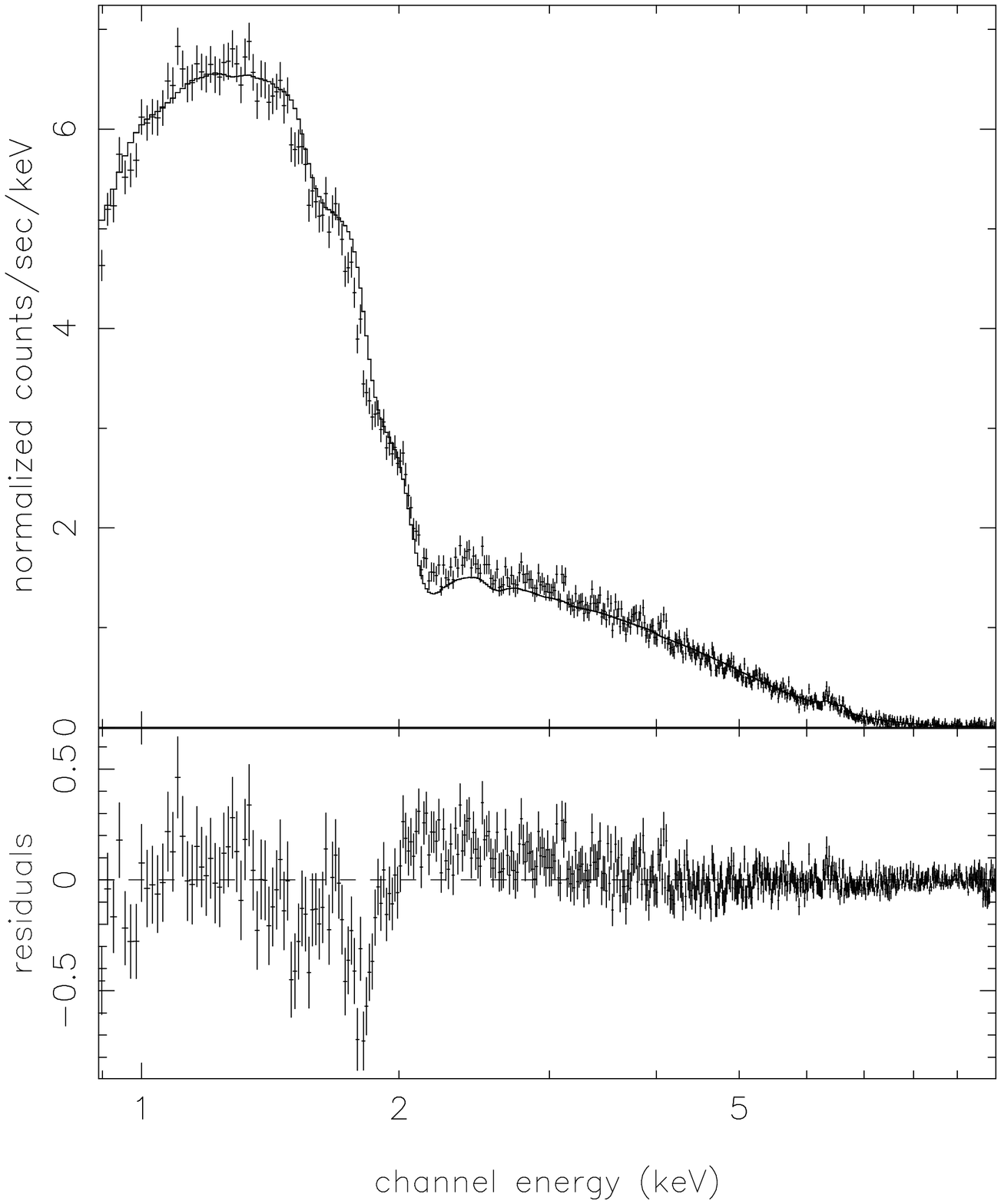}\end{minipage}
   \end{center}}
   {\myputfigure{allspectrum.pdf}{0.0}{1.0}{-40}{-60}}
   \figcaption{\label{fig:allspectrum} 
   Entire cluster spectrum (excluding point sources) and residuals,
	plotted with the
	best-fit MEKAL spectrum described in the text.
     }
\end{inlinefigure}

\begin{figure*}[b]
   \ifthenelse{\equal{\figtype}{EPS}}{
   \begin{center}
   \epsfxsize=16.cm
   \begin{minipage}{\epsfxsize}\epsffile{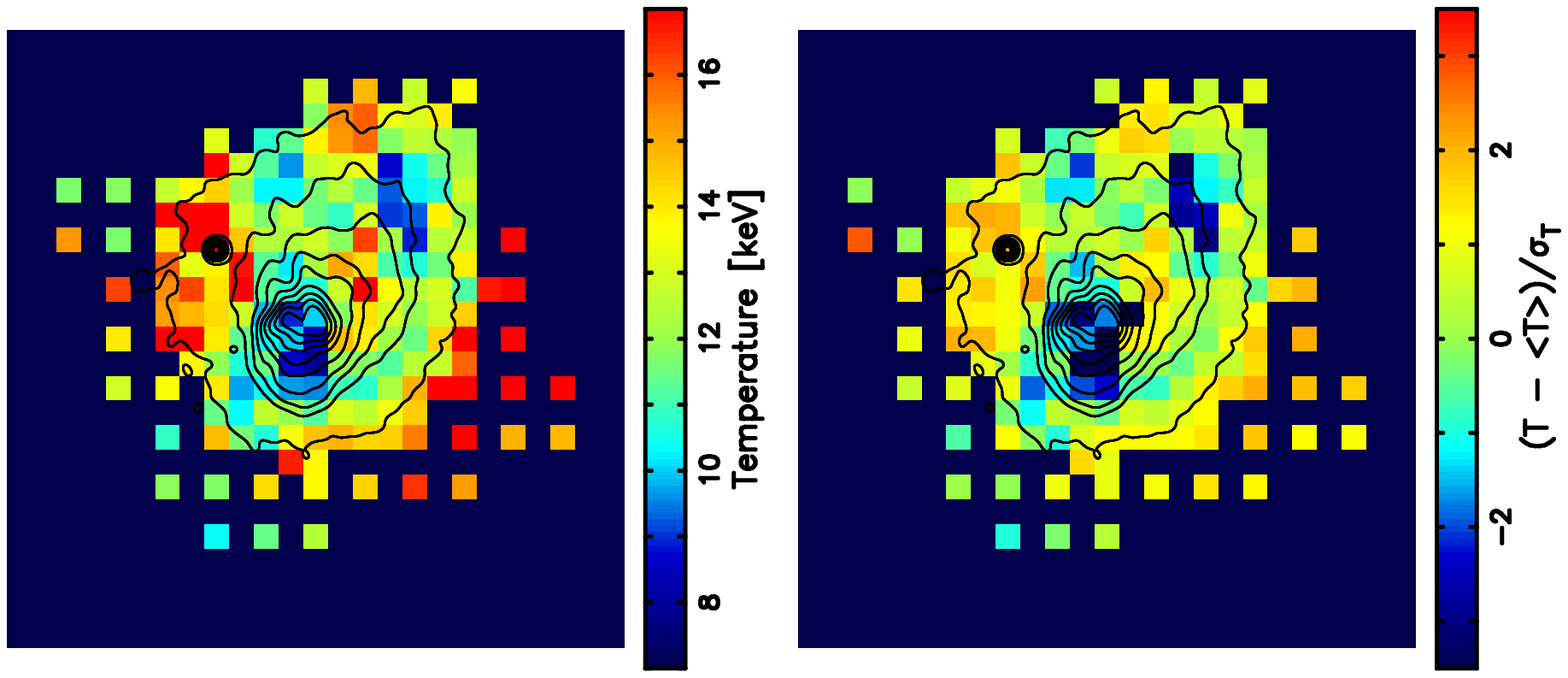}\end{minipage}
   \end{center}}
   {
\myputfigure{tempmaps.pdf}{7.0}{1.15}{-280}{-280}
}
   \figcaption{\label{fig:tempmap} 
   X-ray temperature map ({\it left)}, and significance map ({\it right}).
	The contours
	are from the 0.5--5 keV energy band image shown in 
	Figure~\ref{fig:counts} after smoothing with a Gaussian of constant
	size, and are spaced at 10\% of
	peak cluster intensity. Temperature pixels are $1\arcmin$
	(66 kpc) on a side. The average temperature $\left<T\right> = 11.8 \kev$, and
	uncertainties in this average temperature are not included in the
	significance map.
     }
\end{figure*}

The uncertainty of 0.2 keV given for the cluster temperature above includes only the
statistical uncertainty from the spectral fit. 
We adopt a 1-$\sigma$ uncertainty in \NH\ of $\sim10^{20}$ cm$^{-2}$, which introduces a corresponding 0.5~keV uncertainty in the temperature.  The background subtraction also affects the temperature. The Poisson uncertainty in the background scale factor determined using the 
7--10~keV S2 spectrum is $\sim4\%$, corresponding to a 0.2 keV uncertainty in the cluster temperature.  In addition, the background scaling using the 7--10 keV band produces a cluster temperature that is 0.3~keV higher than that when scaling the background using
the 10--12 keV band;  thus, we adopt a temperature uncertainty contribution from the background scaling of 0.3~keV.
Combining our three sources of uncertainty (statistical, \NH, and background scaling), 
we arrive at a cluster temperature and uncertainty of 11.8$\pm$0.6 keV. 
It should be noted that hydrogen column density 
uncertainties are not included in temperature uncertainties in the rest of the paper unless explicitly noted.

Because X-ray point sources are visible in the \Chandra\ data that were not
noticeable in previous observations, it is possible that their
presence could have affected previous temperature measurements. To check this,
we also fit the entire cluster spectrum without removing point sources; this
produces a temperature decrease of less than 0.1 \kev.

Our measured value of $T_{\rm X} = 11.8 \pm 0.6$ keV is somewhat higher than
previous temperature measurements; our 
abundance value of $0.19 \pm 0.03$ is  low in comparison to previous studies, 
though abundances in this range
appear to be typical in studies of merging clusters \citep{degrandi01}.
The discrepancy between our temperature measurement and previously published
temperatures may be partially explained by \Chandra's relatively small 
field of view and the large angular extension of A2319. 
As is clear from Figure~\ref{fig:rosat}, there is significant cluster emission
outside of the ACIS-I field. Using the PSPC image, we found that $\sim30\%$ 
of the cluster emission in the 0.5--2.0 keV band lies outside of our ACIS-I
observation. A MEKAL model fit on the S2 chip (excluding point sources) 
gives a temperature of $7.1\pm1.2 \kev$ ($\chi^2 = 214$ for 204 d.o.f.). 
This value is in agreement with {\it ASCA} 
measurements of 6--9 keV in large regions around and including 
the area covered by our S2 observation \citep{markevitch96}. 
 If the bulk of the gas outside the
ACIS-I field is similarly cooler than our measured
average temperature of the cluster,
then our temperature measurement with \Chandra\ would naturally be higher than measurements with previous--generation, larger field of view instruments.
This effect probably does not account for the entire difference between our 
result and others, because measurement of temperatures within 
small regions of the ACIS-I chip give slightly higher-than-expected results as
well, as will be discussed in \S\ref{sec:temp}.

If a higher value for \NH\ were used, as discussed above, our fit temperature would be lower.
This cannot account for the discrepancy between our results and previously published 
measurements, however, as previous studies have used column densities within
$\sim0.5\times10^{20}$ cm$^{-2}$ of our adopted value.

The cluster temperature fit is sensitive to the choice of energy band.
For example, fitting the entire cluster spectrum (with abundance and
hydrogen column density allowed to vary)
in the 0.9--10.0 keV band gives $T_{\rm X} = 10.6 \pm 0.3 \kev$ 
($\chi^2 = 999$ for 593 d.o.f.); however, fitting between 1.7--10.0 keV
gives $T_{\rm X} = 6.2 \pm 0.2 \kev$ ($\chi^2 = 705$ for 559 d.o.f.), 
and fitting between 2.0--10.0 keV gives
$T_{\rm X} = 7.6 \pm 0.4 \kev$ ($\chi^2 = 576$ for 523 d.o.f.). 
While the specific
behavior will vary by instrument, it should be noted that the lower
energy limit of most previous temperature measurements has 
been $\sim$1.5--2.0 keV. One obvious explanation for the extreme dependence
of spectral fitting on energy band is simply that the cluster is not 
isothermal, as we show in \S\ref{sec:temp}.

\subsection{Temperature Structure}
\label{sec:temp}

\Chandra\ provides the means to perform a much more detailed study
of the temperature structure of A2319 than previous instruments, permitting
inspection of the cluster merger features. To this end we have made 
an X-ray temperature map of A2319 using the ACIS-I data. 
The map was created by measuring the
temperature at each point on a grid, using a circular
region enlarged until it
contained 2000 counts in the 0.9--2.0 keV energy range. The regions overlap
slightly at the center, and increasingly towards the edge; hence the pixels 
are not independent of one another.
In the faint regions of the observation, where fitting 
region radii are larger than two pixel widths, only
one pixel in four is measured.
The spectra at each point were calculated
using the same procedure as for the whole cluster spectrum described in 
\S\ref{sec:spec}, with abundance floating and 
$N_{\rm H} = 8.33\times10^{20}$ cm$^{-2}$. The abundance was left as a free 
parameter as abundances are known to vary in merging systems; fixing it
to the cluster average produces temperature changes of $< 1 \sigma$
across the temperature map.

\vskip20pt
\begin{inlinefigure}
   \ifthenelse{\equal{\figtype}{EPS}}{
   \begin{center}
   \epsfxsize=8.cm
   \begin{minipage}{\epsfxsize}\epsffile{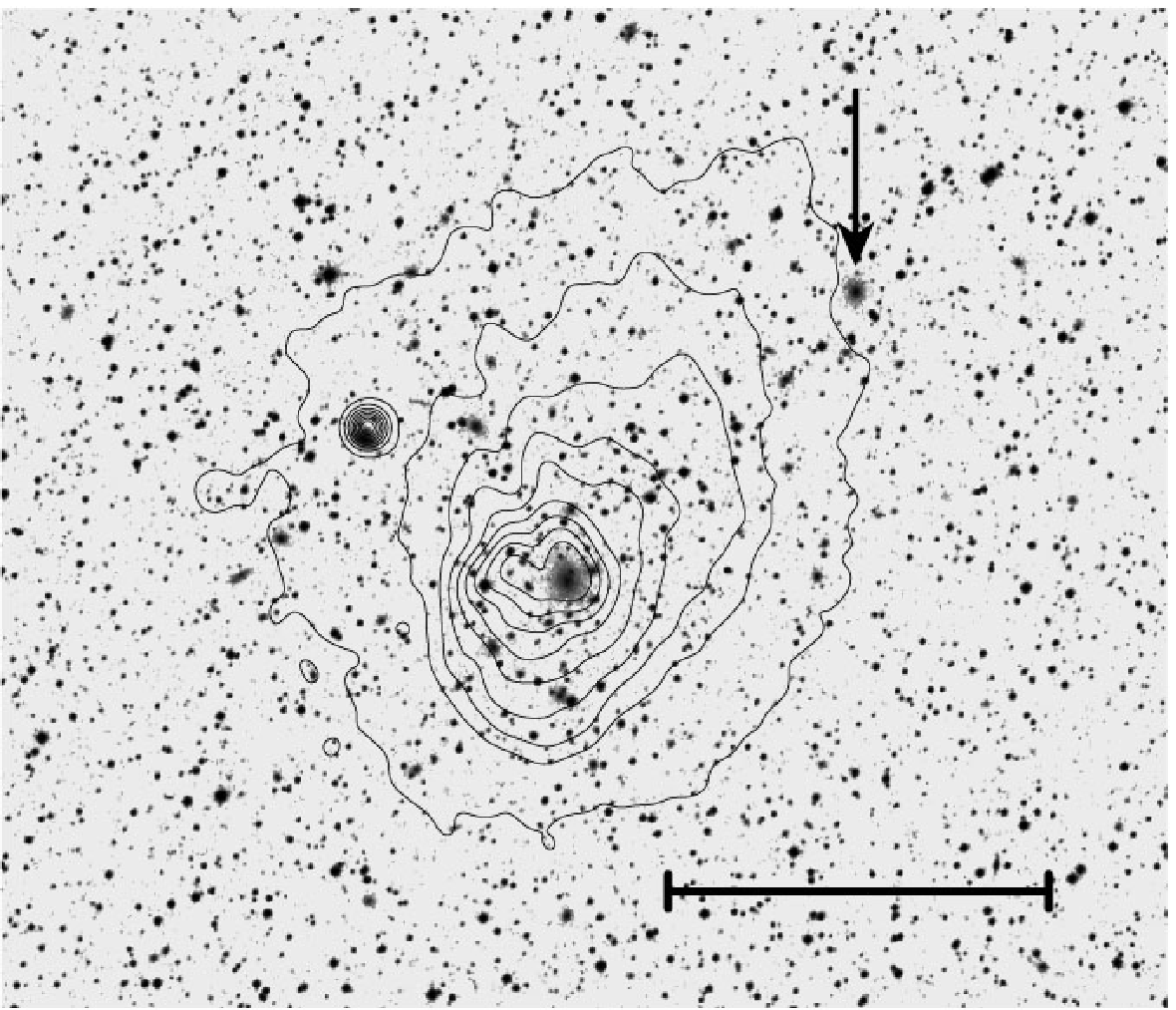}\end{minipage}
   \end{center}}
   {\myputfigure{optical.pdf}{2.6	}{1.7}{-10}{-360}}
   \figcaption{\label{fig:optical} 
   	V-band image from the Digitized Sky Survey 
	with \Chandra\ observation overlaid. Contours are
	the same as in Figure~\ref{fig:tempmap}. The central (i.e., brightest)
	galaxy of A2319B is indicated with an arrow. The bar indicates 
	a distance of 0.5 Mpc.
     }
\end{inlinefigure}

The temperature map is shown in Figure~\ref{fig:tempmap} (left) with 
overlaid surface brightness contours.
Also in Figure~\ref{fig:tempmap} (right) is a map of the significance of 
deviations from the mean temperature; that is, the difference between
each pixel temperature and our adopted cluster mean temperature of 11.8 keV,
divided by the uncertainty in the pixel temperature.  The general structure 
of the temperature map includes two cooler-than-average regions that lie 
along a northwest--southeast line, and 
possibly two hotter-than-average regions that 
lie to the northeast and southwest of center.  This temperature morphology 
is suggestive of a merger along a northwest--southeast trajectory, where 
remnants of cold cores remain and shock heated gas is escaping perpendicular 
to this merger axis, as seen in hydrodynamical simulations 
\citep{roettiger97,ricker01}.
The cold spots deviate from the mean by $>2\sigma$; the hot spots are somewhat
less significant.
The very high ($\gtrsim15 \kev$) temperature regions lie where the cluster 
surface brightness is lowest, making these temperatures particularly 
susceptible to background subtraction errors.
Overall, temperatures are higher than would be expected based on previous
studies of A2319 \citep{markevitch96}. Regardless of any overall temperature increase, the
nonisothermality of the cluster provides some indication as to the origin
of the poor fit discussed in \S\ref{sec:spec}.

The level of substructure revealed here is more detailed than has been 
previously seen. The coolest region lies just south of the surface 
brightness peak, perhaps indicating a cool core that has thus far survived
the ongoing merger. It is not immediately obvious from this map 
whether there is a 
sharp temperature change across the merger feature significant enough to
deduce the existence of either a shock front or a cold front;  we examine this in more detail in \S\ref{sec:mergprofile}.

This cool core has not been identified in the earlier \ASCA\ temperature map \citep{markevitch96}. 
It seems likely that surrounding areas of higher-than-average temperatures obscured the core
in the lower angular resolution \ASCA\ map. 
\citet{molendi99} pointed out a ``subcluster'' of temperature 
$6.9\pm1.0 \kev$ to the northwest of the cluster center, and there is evidence
for the presence of this cool region in the temperature map of 
\citet{markevitch96}. 
This subcluster is seen here $\sim6\arcmin$ northwest
of the X-ray brightness peak, though at a somewhat higher temperature. 
Also present is a distinct region of somewhat elevated (i.e., 
above the mean) temperatures between this subcluster and the cool center.

The cool ICM ``subcluster'' has been identified with subcluster A2319B; 
however, at this
resolution it is clear that the cool region is not associated with the
center of A2319B, but rather lies 2--3$\arcmin$ to the east-southeast of it,
as can be seen by comparing
the temperature map to the visual-band image shown in Figure~\ref{fig:optical}.
This suggests that the subcluster is in a transient phase wherein the 
ICM has decoupled from the galaxies. Such a state has been
observed in other merging clusters such as 
1E0657-56 \citep{markevitch02}
A754 \citep{zabludoff95, markevitch03},
Cl J0152.7--1357 \citep{maughan03},
and A2034 \citep{kempner03}.

Overall, the temperature map reveals complex substructure of the type now
known to occur in galaxy cluster mergers. Such substructure is also seen in
hydrodynamical simulations \citep[e.g.][]{roettiger97,onuora03}

\section{Merger Analysis}
\label{sec:merger}

We present here a simple analysis of the
merger features in A2319, wherein we assume a simple spheroidal geometry
for an isothermal body of gas
falling into a relaxed $\beta$-model cluster. This is what
might be termed the ``traditional'' analysis of a merging cluster \citep[following][]{vikhlinin01}.  
However, numerical simulations of clusters \citep[e.g., ][]{ricker01,bialek02,nagai03,onuora03}
have made it clear that the dynamics within
a mid-stage merger are much more complex than this. Nevertheless, this
naive analysis offers some level of quantitative information about the
nature of the merger front, and permits comparison to other merger analyses.

\subsection{Temperature and Brightness Profiles Across Merger Feature}
\label{sec:mergprofile}

We measure the surface brightness and temperature
profiles across the merger feature (see Figure~\ref{fig:radprofile}). The brightness is measured in arcs on a wedge, chosen with a radius of curvature
and angular width that match the brightness
discontinuity reasonably well. We then measure
the temperature, making spectra as previously described, in arc segments
of sizes chosen both to provide a sufficient number of photons and to
permit study of temperature variation across the front; we select the segment boundaries  to avoid
having a region straddling the surface brightness discontinuity.   Note that this is not a cluster radial
profile; the wedge in which this is measured is chosen to match the 
brightness discontinuity, and is not centered on the brightness peak.

While there is clearly a brightness change, this change is not as sharp as those seen in merging clusters such as A3667 \citep{vikhlinin01}.  This can be readily
explained if the merger is not taking place close to perpendicular to the
line of sight; indeed, the aforementioned difference in line-of-sight velocity
between A2319A and A2319B of $\sim2900$ km s$^{-1}$ \citep{oegerle95}
suggests that we are viewing the merger at some
large angle. This introduces substantial uncertainties
into the analysis below.

\vskip20pt
\begin{inlinefigure}
   \ifthenelse{\equal{\figtype}{EPS}}{
   \begin{center}
   \epsfxsize=8.cm
   \begin{minipage}{\epsfxsize}\epsffile{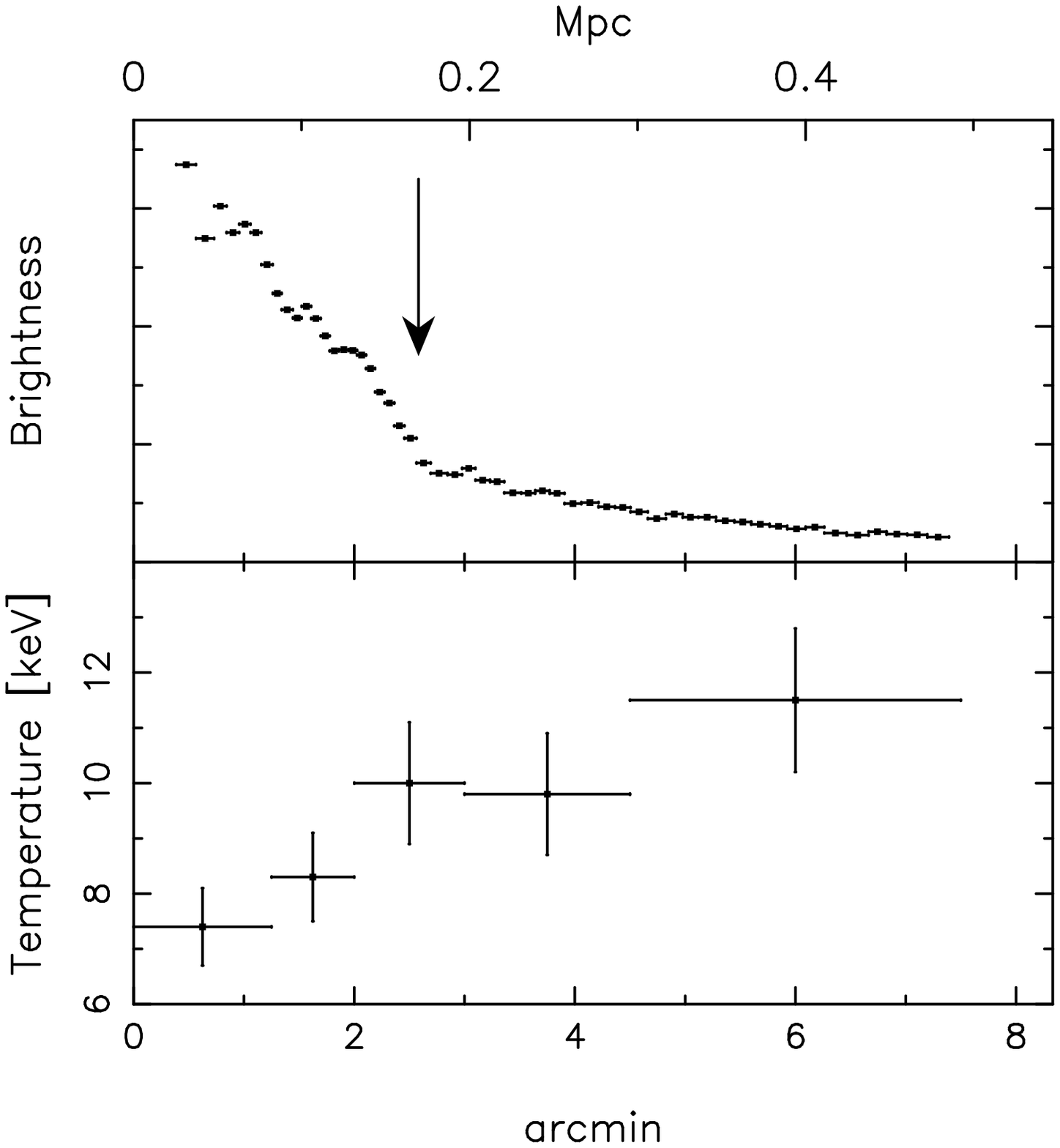}\end{minipage}
   \end{center}}
   {\myputfigure{mergerprofile.pdf}{0.0}{1.2}{-60}{-110}}
   \figcaption{\label{fig:radprofile} 
   Brightness (in arbitrary units)
	and temperature profiles across the merger feature. The arrow
	indicates the approximate position of the brightness discontinuity
	seen in Figure~\ref{fig:counts}. A fit to the surface brightness inside the
	merger feature using Eq.\ref{eq:frontbrightness2} is shown by the solid line.
     }
\end{inlinefigure}

If we assume that the infalling gas body is a spheroid with constant gas
density, then the surface brightness profile at distances from the front 
much less than the major axis of the spheroid will be given by

\begin{equation}
\label{eq:frontbrightness1}
S(d) = 2^{3/2} \sqrt{R} \varepsilon_0 \sqrt{d}~,
\end{equation}

\noindent where $R$ is the radius of curvature at the front, $\varepsilon_0$ is the 
volume emissivity of the gas, and $d$ is the distance from the front
\citep{vikhlinin01}. This function adequately describes the surface 
brightness profile of our data in the region just inside the front 
(i.e., the region between roughly 1.5--2.8$\arcmin$ 
in Figure~\ref{fig:radprofile}).

More precisely, the surface brightness profile is

\begin{equation}
\label{eq:frontbrightness2}
S(d) \simeq 2a\varepsilon_0\left(\frac{2d}{b}-\frac{d^2}{b^2}\right)^{1/2}
	\left(1-\frac{d}{b}\right)^{-3.45\beta}~
\end{equation}

\noindent for $|\beta| < 0.25$, where $a$ and $b$ are the short and long
axes of the spheroid, respectively \citep{vikhlinin01}.
Fitting this function to the brightness profile just interior to the front
gives $\beta\lesssim 0.1$. Our approximation of constant density inside
the front is thus justified.

An examination of Figure~\ref{fig:radprofile} does not conclusively determine
the nature of the merger feature, i.e., whether it is a shock front 
or cold front. The temperature just inside the merger is $9.0\pm0.9 \kev$, 
while the temperature just outside is $10.8 \pm 1.3 \kev$. This suggests
that the feature is a cold front, but the temperature uncertainties are large. The temperature falls
by another 1--2 keV deeper inside the infalling subcluster, perhaps suggesting some heating near the surface brightness discontinuity.

To rule out significant upward biasing of temperatures inside the front by
projected hotter gas in front of and behind the cooler gas, we fit a 
two-component MEKAL model to a region inside the front, near the 
brightness discontinuity, with the hotter component fixed to the temperature
measured outside the discontinuity. We find that to measure a cool component
temperature that is 1 $\sigma$ lower than the single-component temperature 
measurement
requires a hot component contribution of $\gtrsim 40\%$ of the emission.
 As this
seems unreasonably high, we conclude that our temperature measurements inside
the front are not significantly biased by projected hotter gas.

\subsection{Density Variation Across Merger Feature}
\label{sec:mergtemp}

In general, the intensity
 of a body of gas at constant temperature is

\begin{equation}
\label{eq:intensity}
I = \frac{1}{(1+z)^4}\int n_e n_H \Lambda(T_{\rm X},l) \, dl~,
\end{equation}

\noindent where $\Lambda(T_{\rm X},l)$ is the emissivity of the gas
and the length element $dl$ is along the line of sight; 
the integration is carried out over
the entire body along the line of sight.

If the spheroid's long axis is much larger than the minor axes, we
can model the infalling subcluster as a ``bullet'' of width $L$; we assume
a constant temperature and intensity. Using these assumptions in 
equation~(\ref{eq:intensity}) and solving for the electron density gives

\begin{equation}
\label{eq:bulletdensity}
n_e = \left(\frac{I}{L\Lambda(T_{\rm X})}\frac{\mu_e}{\mu_H}\right)^{1/2}(1+z)^2~.
\end{equation}

\noindent We use $\mu_e = 1.67$ and $\mu_H = 1.4$,
the values for a fully ionized gas of one-third solar abundance. Using
estimated values for $I$ and $L$, we obtain an electron density immediately
inside the merger front of $(6.0 \pm 1.0)\times10^{-3}$ cm$^{-3}$. 

To get the electron density outside the front, we assume that the gas
fits a spherical $\beta$-model density profile:

\begin{equation}
\label{eq:betamod}
n_e = n_{e_0} \left(1+\left(\frac{\theta}{\theta_c}\right)^2\right)^{-3\beta/2}~,
\end{equation}

\noindent with central electron density $n_{e_0}$ and critical radius 
$\theta_c$. 
That is, we assume that the gas to the southeast of the front is part of the
original relaxed cluster into which a subcluster is falling, and is thus far
unperturbed by the merger. To get values for $\beta$ and $\theta_c$, 
we fit the surface brightness in our wedge, outside the merger. We arrive at
an electron density immediately outside the merger feature of 
$(2.0 \pm 0.6) \times 10^{-3} $ cm$^{-3}$, or roughly one-third the density
immediately inside the feature. 

These densities correspond to pressures 
($p = n_e T_{\rm X}$) inside of 
$p_{\rm in} = (5.4\pm1.0) \times 10^{-2}$ keV cm$^{-3}$
and outside of $p_{\rm out} = (2.2\pm0.7)\times 10^{-2}$ keV cm$^{-3}$.
Using the relationships between these pressures and the Mach
number $M$ of the infalling gas cloud (where $M = v/c_{\rm out}$ is the
Mach number in the free stream outside the merger) gives $M = 1.1 \pm 0.3$ \citep[see \S122;][]{landau87}. The infalling cluster 
would thus probably be moving at a roughly sonic speed if indeed the merger
were taking place in the plane of the sky, as we have assumed for this 
analysis. Because of the line of sight velocity difference of the galaxies associated with A2319A and A2319B, we expect that the merger axis does not lie in the plane of the sky.

\subsection{Toward a Cluster Dynamical Model}
\label{sec:mergdisc}

Combining the results of the previous sections, we present the following
picture of the merger in A2319. There is a jump by a factor of 
$3.0 \pm 1.0$ 
in the density of the gas as one crosses the brightness discontinuity
from the unperturbed gas outside the merger towards the cluster core. This
is accompanied by a slight temperature decrease of 1--3 keV, and a brightness
increase by a factor of $\sim 3$; the combined densities and temperatures 
give a pressure jump by a factor of $2.5 \pm 0.9$. These results indicate
the presence of a cold front, although the temperature difference across the front is not
as large as is observed in, e.g., A3667 \citep{vikhlinin01}.

However, we have assumed for this analysis that the infalling subcluster 
is moving
in the plane of the sky; our value for the electron density in the cool core
is thus an overestimate given the known line-of-sight velocity
difference between A2319A and A2319B that indicates that bulk gas motions
are not perpendicular to the line of sight. 
This is most easily seen by examining 
equation~\ref{eq:bulletdensity}; if the subcluster is not oriented
perpendicular to the line of sight, then we are overestimating the
X-ray intensity $I$, and hence also the electron density $n_e$. Moreover, if
this merger has a nonzero impact parameter,
then our estimate for the ambient electron density, i.e., the density outside
the merger feature, is likewise an overestimate.
It is thus possible that the inside/outside density and 
pressure ratios
are in fact lower or, less likely, higher than the values given. This does
not change the general interpretation of the merger feature as a cold front;
the temperature change is indisputable, and the uncertainties in density
are not large enough to accomodate a pressure outside the front greater
than that inside the front.

The simplest interpretation for the merger geometry seen in A2319 is that
A2319B has recently fallen through A2319A, in the process losing much of
its ICM as indicated by the low X-ray brightness around the giant elliptical that dominates its galaxy population. The
separation of the cold spot near A2319B from its galaxies supports this  (see Figures~\ref{fig:tempmap} \& \ref{fig:optical}). 
However,  the structure of the cold front suggests motion away from A2319B. We suggest that the encounter of the
two subclusters of A2319 has caused the cool core of A2319A to be displaced
from its position at the center of the subcluster, and that this core
is now recoiling from that displacement and has passed its original, central
position. This is supported by the fact that the coldest part of A2319A's core
is located slightly to the southeast of the brightest cluster galaxy. The
merger feature is then a result of the interaction of the dense core ICM with
less dense, warmer ICM surrounding the core.

The apparent survival in some form of the cold core ICM of A2319B may indicate
a non-zero impact parameter. Given this and the relative sizes of the two
subclusters, displacement of the core of A2319A to the point of creating
motion of the core at near-sonic speeds would require a large infall velocity.
It is also possible that the merger was essentially head--on, and that cool ICM spatially associated with the galaxies of A2319B is not from the original core of the subcluster, but
was pulled from the core ICM of A2319A during the collision \citep[see][]{pearce94}.

We estimate the timescale since closest approach of the two subclusters by constructing a simple, two-body dynamical model.  Using the line of sight velocity dispersion of A2319A
\citep[$\sigma_A = 1324$ km s$^{-1}$,][]{oegerle95}, we obtain a crude estimate of the collision infall
velocity of $\sqrt{6}\sigma_A = 3243$ km s$^{-1}$ (this assumes infall from infinity). Combined with the
measured line-of-sight velocity difference of subclusters A and B  \citep[2909 km s$^{-1}$][]{oegerle95},
we estimate that the merger trajectory has an angle of $\sim 65^\circ$ out of the plane of the sky.  The corresponding velocity in the plane of the sky is $\sim 1430$ km s$^{-1}$. This gives
a time since closest approach of the subcluster cores of $\sim 0.4$ Gyr.

We emphasize that this is only one possible merger scenario. It does
not include the possibility of a second merger event taking place along
the northeast-southwest direction such as that suggested 
by an analysis of earlier X-ray data \citep{feretti97}.

\section{Cluster Observables During a Major Merger}
\label{sec:structure}
The merger signatures in Abell~2319 are clear. These include significant 
centroid shifting in the {\it Einstein} IPC \citep{mohr95} 
and \ROSAT\ PSPC X-ray images; two subclusters identified in the optical \citep{faber77,oegerle95}; 
differing distributions of galaxies and ICM; and a surface brightness 
discontinuity and temperature structure in the \Chandra\ data.
We seek now to examine how mergers perturb the global physical structure of
clusters.
We address this empirically by simply 
examining how particular bulk properties of A2319 (binding mass, ICM mass, 
isophotal size, luminosity, emission weighted mean temperature, and galaxy light) 
compare to typical galaxy clusters.  Specifically, we compare the properties 
of A2319 to what is essentially a flux-limited ensemble of 44 galaxy clusters 
from the nearby universe \citep{mohr99}. While the high resolution of 
\Chandra\ is not necessary for this, the \Chandra\ observation nonetheless
provides another high-quality data set for such study.

The question of how much merging perturbs the global structure of galaxy clusters is particularly important in light of the planned and ongoing high-yield galaxy cluster surveys.  In these surveys, rather simple observables like the SZE flux, X-ray flux, and galaxy light will be used to estimate cluster masses for studies of the dark energy \citep[e.g., ][]{haiman01}.  Even though it has recently been shown that very large surveys contain
 enough information to self--calibrate while precisely constraining the dark energy \citep{majumdar03a,majumdar03b,hu03b}, any improvements in our understanding
 of cluster mass--observable relations, their evolution, and the effects of merging on them will lead to tighter limits on systematic uncertainties in the resulting cosmological constraints.

\vskip20pt
\begin{inlinefigure}
   \ifthenelse{\equal{\figtype}{EPS}}{
   \begin{center}
   \epsfxsize=8.cm
   \begin{minipage}{\epsfxsize}\epsffile{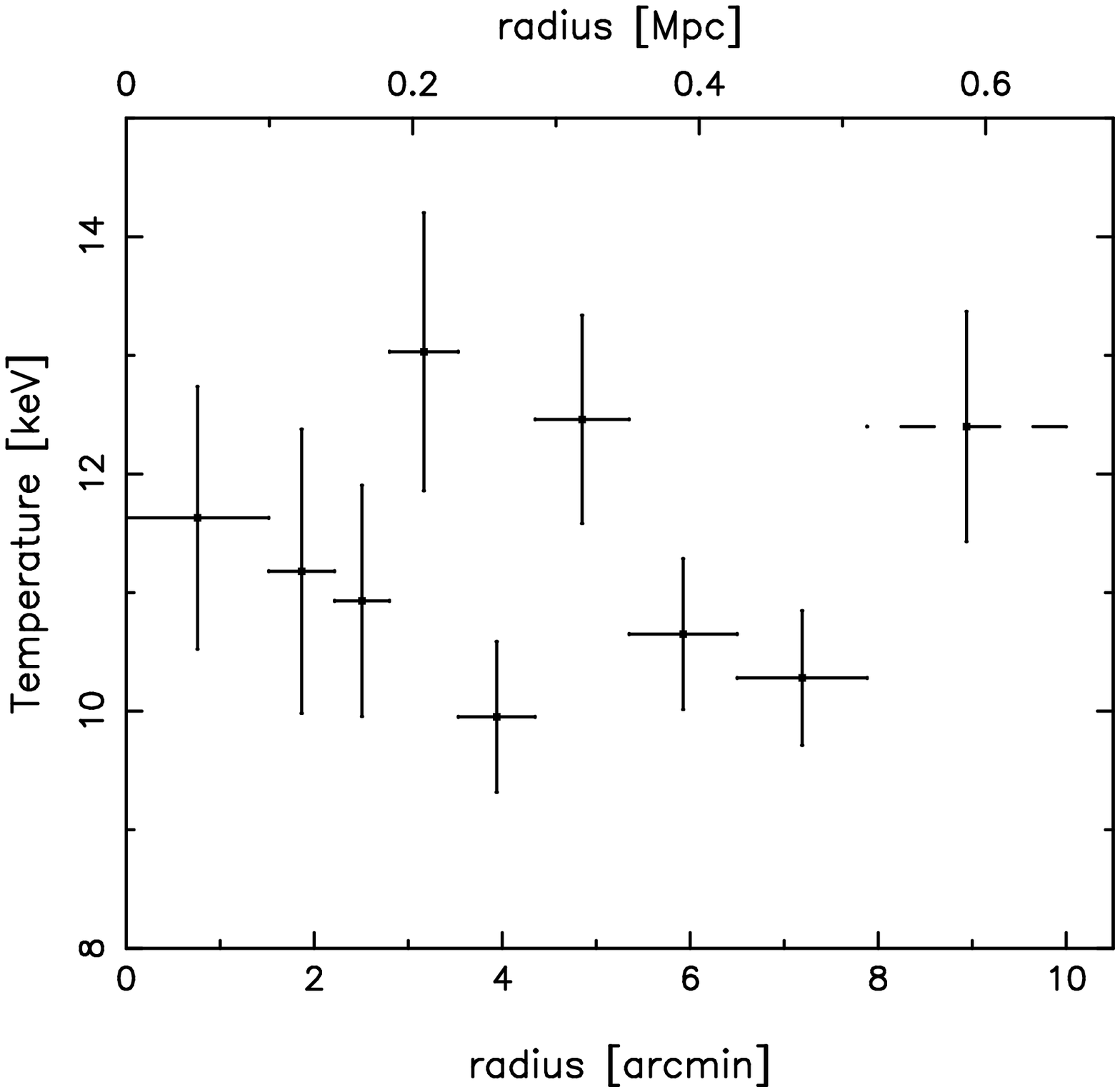}\end{minipage}
   \end{center}}
   {\myputfigure{antempprofile.pdf}{0.0}{1.0}{-80}{-60}}
   \figcaption{\label{fig:annulartemp} 
Projected temperature profile created around the center of the cluster as
	found by a $\beta$-model fit, not around the surface brightness peak. The
	outermost annulus partially extends beyond the boundaries of the 
	ACIS-I observation, and so we mark the 
	angular extent of this annulus with a dashed line. 
     }
\end{inlinefigure}

\subsection{Naive Analysis of Mass Profile}
\label{sec:massprof}

We obtain a naive measure of \M2500, i.e., 
the mass enclosed by \r2500, the
radius within which the mean density is 2500 times the critical density
of the universe. Most cluster surveys focus on properties
at larger radii (e.g., $r_{500}$), but \Chandra's small field of view 
relative to the large angular extent of A2319 makes this impossible with a single pointing.

We assume that the cluster ICM density distribution is fit
by a spherical $\beta$-model and that the ICM is in hydrostatic equilibrium; the binding mass within
a radius $r$ is then
\begin{equation}
\label{eq:massfunc}
  M(r) = -\frac{k_B}{\mu m_p G} r T(r)
	\left(\frac{\partial \ln \rho_{\rm gas}(r)}{\partial \ln r}
	+\frac{\partial \ln T(r)}{\partial \ln r} \right) ~.
\end{equation}
Figure~\ref{fig:annulartemp} contains a projected temperature profile of A2319, where the cluster center is that found by the $\beta$-model fit to the X-ray surface brightness (described below). 
There is no easily quantifiable variation of temperature as a function of
distance from the cluster center. Given the measured density and temperature profile, 
\r2500 lies mostly outside the ACIS-I image.  To estimate the mass at this radius, we adopt an
isothermal temperature profile and extract the average temperature from the outer three annuli in Figure~\ref{fig:annulartemp}.  This temperature is $11.1\pm0.9$~keV, less than $1\sigma$ lower 
than the emission weighted mean temperature for the cluster.

A fit to the \Chandra\ surface brightness image of A2319
gives core radius $r_c = 0.17 \pm 0.01$~Mpc 
($\theta_c = 2\arcmin.6 \pm 0\arcmin.1$) and
$\beta = 0.55 \pm 0.01$ 
\citep[compared to $r_c = 0.15\pm0.05$~Mpc and $\beta=0.54\pm0.06$ from the analysis of the PSPC image;][]{mohr99}. 
With these values and equation~(\ref{eq:massfunc}), we find
$r_{2500} = 0.67\pm0.02$ Mpc  ($\theta_{2500} = 10\arcmin.2 \pm 0\arcmin.4$) 
and binding mass $M_{2500} = (4.2\pm0.5)\times10^{14}~M_\odot$.  
The uncertainties quoted here for the $\beta$-model fit are 1 $\sigma$ 
statistical uncertainties only, and do not reflect the fact that the $\beta$ model is not a particularly
good fit the surface brightness in this complex cluster.  The mass uncertainty
is dominated by the uncertainty in the temperature measurement.

We compare this mass estimate to that expected for a cluster with this emission-weighted mean temperature, using an \M2500-\Tx\ relation derived from a sample of seven intermediate-redshift clusters \citep{allen01}. For our cluster temperature
of $T_{\rm X} = 11.8 \pm 0.6 ~\kev$ the best fit relation gives 
$M_{2500} = (6.7\pm0.8) \times 10^{14} M_\odot$, which is  
a factor of $1.6\pm0.3$  higher
than our value.  
In this merging cluster, the hydrostatic equilibrium assumption and spherical $\beta$-model fitting thus lead to a mass estimate that lies  $\sim60\%$ off the relation found in apparently ``relaxed'' clusters.  The \citet{allen01} sample is too small to make meaningful statements about the scatter, but other analyses of much larger samples show scatter at roughly the 25\% level \citep{finoguenov01}.

\subsection{Comparison of A2319 to Large Cluster Sample}
\label{sec:observables}

We examine five bulk properties of A2319---the X-ray luminosity, emission-weighted mean temperature, ICM mass, isophotal size, and $K$-band galaxy light---and compare these properties to the same properties for large samples of galaxy clusters.  In the case of all but the galaxy light we use an ensemble of 44 clusters studied using the \ROSAT\ PSPC \citep{mohr97a,mohr99,mohr00a}, but reanalyzed at the cluster radius \r2500.  For each cluster we determine \r2500\ using the emission weighted mean temperature and the published \M2500--temperature relation \citep{allen01}.  For the PSPC sample, exposure corrected, background subtracted images were prepared in the rest frame 0.5--2.0~keV band for each cluster.  In the case of the galaxy light, we compare to an ensemble of $\sim$100 clusters whose properties are being studied using X-ray data and 2MASS near-IR data \citep{lin03c}.  

We measure \L2500, the luminosity projected
within a circle of radius \r2500\ in the 0.5--2.0 keV band. Due to the uncertainty
of the spectral response of \Chandra\ below $\sim 0.9$ keV, 
for our observation of A2319
we measure the flux or luminosity in an image that includes only counts in the 0.9--2.0 keV band.
  With the emission-weighted mean temperature, we calculate the conversion between the count rate in this band and the flux within the rest frame 0.5--2.0 keV band.   Another difficulty is that 
we find $\theta_{2500} = 11\arcmin.8$ from the \citet{allen01} \M2500-\Tx\
relation, slightly too large to fit 
within the ACIS-I observation. However, the low luminosity near the edges,
relative to the central luminosity, means that our value 
$L_{2500} = 5.2 \times 10^{44}$ erg s$^{-1}$ contains the bulk of \L2500.
Indeed, using the \Chandra\ footprint on the \ROSAT\ PSPC image, we find that 
15\% of the flux within \r2500\ is missed; 
thus, our corrected estimate of the luminosity in the 0.5--2.0~keV band is
$L_{2500} = 6.0 \times 10^{44}$ erg s$^{-1}$.  This is high by $\sim$18\% compared to the value $L_{2500} = 5.1 \times 10^{44}$ erg s$^{-1}$ measured using the PSPC image.   Figure~\ref{fig:panel} (top) contains the \ROSAT\ sample (small points) with best fit power law together with the \Chandra\ measurement (large point).  The luminosity is low by 54\% relative to the expectation for a cluster with a temperature of 11.8~keV, compared to an RMS fractional scatter about the best fit relation of 57\% (the PSPC value for \L2500\ is low by $\sim61\%$).

\begin{inlinefigure}
   \ifthenelse{\equal{\figtype}{EPS}}{
   \begin{center}
   \epsfxsize=8.cm
   \begin{minipage}{\epsfxsize}\epsffile{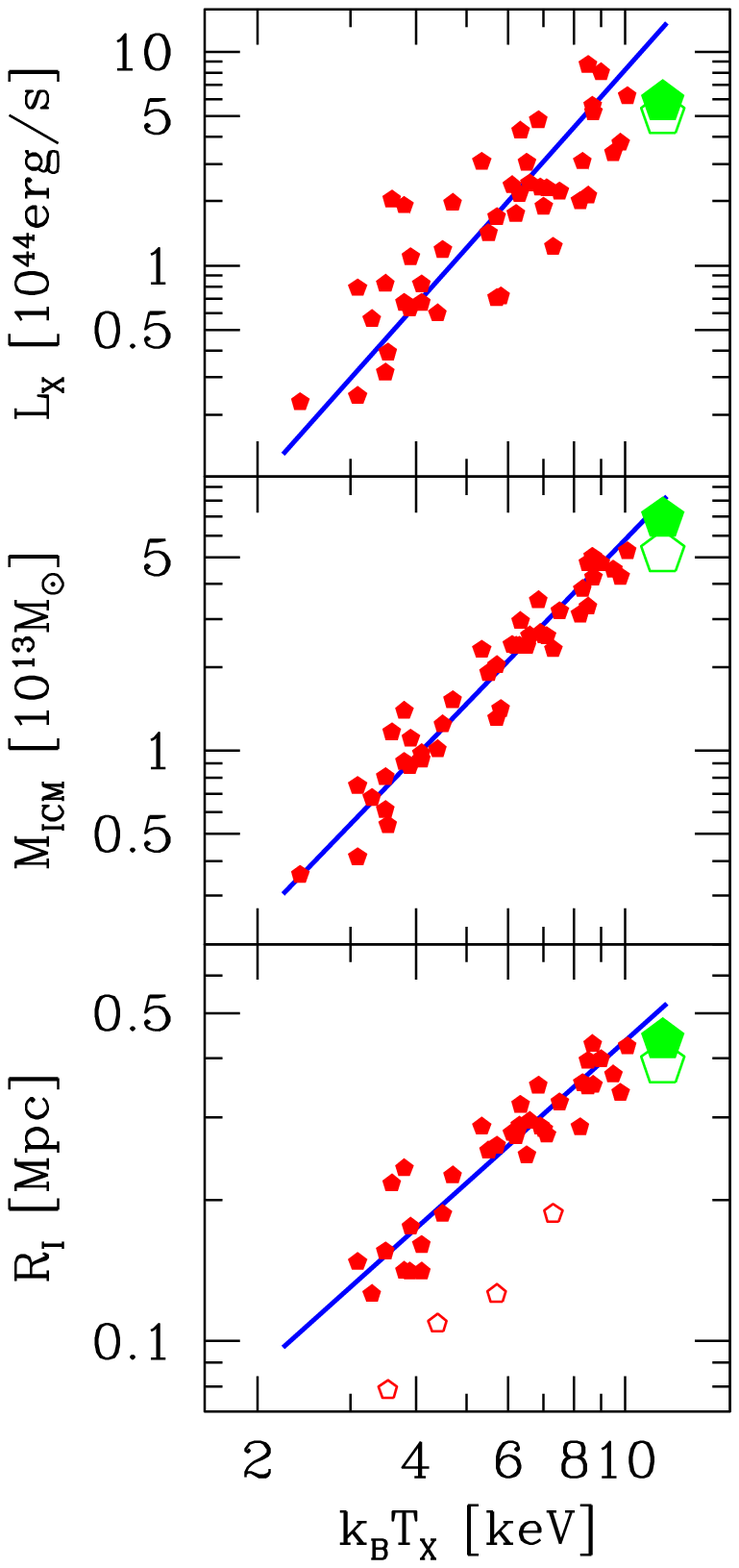}\end{minipage}
   \end{center}}
   {\myputfigure{panel.pdf}{4.0}{2.2}{-210}{-90}}
   \figcaption{\label{fig:panel} 
   Scaling relations for the sample of \ROSAT\ PSPC observations from
	\citet{mohr99} (small points), plus our measurements for A2319
	({\it Chandra} large, solid point; PSPC large, open point), using a temperature
	 of 11.8 keV. Best fits to the PSPC
	sample are shown as lines.
	{\it Top}: \L2500\ in the 0.5--2.0 keV band.
	{\it Middle}: \Micm\ within \r2500.
	{\it Bottom}: Isophotal radius for an isophote of 
	  $1.53\times10^{-13}$ erg s$^{-1}$ cm$^{-2}$ arcmin$^{-2}$ in
	  the 0.5--2.0 keV band. Points shown in outline were excluded from
	  the fit, as the use of a very high isophote caused them to give
	  erroneous results.
     }
\end{inlinefigure}

We measure the ICM mass within \r2500\ using a measurement of the flux from the cluster combined with the $\beta$-model fit parameters and our cluster temperature of $11.8 ~\kev$.  The count rate emissivity of a parcel of gas within the 0.5--2.0~keV band has low sensitivity to temperature variations and, assuming all the ICM is emitting at the emission weighted mean temperature, provides a good estimate of the ICM mass \citep[see][]{fabricant80,mohr99}.
The ICM mass from the \Chandra\ analysis is
$M_{\rm ICM} = 6.9 \times 10^{13} M_\odot$, corresponding to an ICM mass fraction
of $f_{\rm ICM} = 16\%$.  The corresponding value from the PSPC analysis is $M_{\rm ICM} = 5.2\times 10^{13} M_\odot$, roughly 30\% lower.  
Figure~\ref{fig:panel} (middle) contains the \ROSAT\ sample (small points) with best fit power law together with the \Chandra\ measurement (large point).  The ICM mass is low by 
$\sim14$\% relative to the expectation for a cluster with a temperature of 11.8~keV, compared to an RMS fractional scatter about the best fit relation of 20\% (the PSPC value for the
ICM mass is low by $\sim35\%$).

We measure the isophotal size \RI\ for A2319 at an isophote of
$1.53 \times 10^{-13}$ erg s$^{-1}$ cm$^{-2}$ arcmin$^{-2}$ 
in the 0.5--2.0 keV band. This isophote is chosen so that the isophotal size is not affected by the limited field of view of the \Chandra\ footprint.  We measure the size using  the area $A_{\rm I}$ enclosed by 
this isophote, and find an equivalent radius from 
$A_{\rm I} = \pi R_{\rm I}^2$.  
For the \Chandra\ observation we obtain $R_{\rm I} = 0.44$ Mpc, 
compared to the value $R_{\rm I} = 0.39$ Mpc obtained when using the PSPC image.
Figure~\ref{fig:panel} (bottom) contains the \ROSAT\ sample (small points) with best-fit power law together with the \Chandra\ measurement (large point).   
Note that at an isophote this bright
there are several clusters that simply fall off the relation defined by the bulk of the PSPC clusters.  We are forced to use such a bright isophote because the \Chandra\ footprint is so small compared to the angular extent of A2319.  Nevertheless, A2319's isophotal size is $\sim14$\% lower than that expected for a cluster with its emission weighted mean temperature, compared to an RMS fractional scatter about the best fit relation of  14\% (the PSPC value for 
$R_{\rm I}$ is low by $\sim24\%$). 

Interestingly, A2319 does not stand out significantly from the sample of 44 clusters (essentially an X-ray flux-limited sample) studied with the \ROSAT\ PSPC.  The merger in A2319 does not perturb the cluster significantly in luminosity, ICM mass, or isophotal size from the values expected for a cluster with its emission weighted mean temperature.  
In addition, an analysis of the galaxy light in the $K$-band that is projected within \rfive\ in A2319 leads to an estimate of the cluster $K$-band light that is 14\% higher than expected for a cluster with a 11.8~keV temperature, when compared to a sample of $\sim100$ clusters where the rms scatter is 30\% \citep{lin03c}.

One possible explanation is that the merger event is relatively minor (the ratio of velocity dispersions of  A2319A and A2319B suggests a mass ratio of $\sim$8), but it may also be that merging clusters are perturbed in all their quantities in such a way that they remain close to the population-defined scaling relations.  In fact, it should be noted that many of the clusters contained in the PSPC sample exhibit evidence for ongoing mergers \citep{mohr95,buote96}.
We cannot hope to deliver a final verdict on the effects of merging on the bulk properties of clusters with 
studies of single clusters; however, our results do provide some evidence
that bulk properties either do not change much as a result of mergers, or change in a correlated way that maintains the strikingly small scatter of scaling relations.

Correlated changes in luminosity and temperature within merging clusters have been examined with numerical simulations. \citet{ricker01} measured
the luminosity and temperature boosts in merging cluster systems as a function
of time. If we assume a 1:3 mass ratio for the subclusters,
then the simulations predict a peak luminosity boost by a 
factor of $\sim$ 2--4, along with a peak temperature boost of a factor of $\sim$ 1.5--2.0,
with correspondingly smaller boosts associated with larger mass ratio mergers.
Simultaneous boosts to the luminosity and temperature of these magnitudes
would not make A2319 stand out in the luminosity--temperature relation in
Figure~\ref{fig:panel}.  However, A2319 appears ``normal'' with respect to its luminosity--temperature, isophotal size--temperature, ICM mass--temperature and galaxy light--temperature properties.  It would seem to be contrived to claim that large, merger related excursions in these five bulk properties of the cluster all take place in just such a way as to keep the cluster near the observed, typical behavior for a large sample of clusters.  A simpler explanation would appear to be that these five cluster properties are simply not dramatically affected by the merger taking place in A2319.

\section{Conclusions}
\label{sec:concl}

Using \Chandra\ data, we have identified and studied a major merger event
in A2319 that appears to be taking place along the axis connecting its two major optical subclusters.  The X-ray
brightness map shows a clear discontinuity that appears similar to cold fronts found in other clusters.  Although this cold front appears to be as large as the one studied in A3667 \citep{vikhlinin02}, it is not as sharp. This, together with previous measurements of the line of sight velocity difference between the two main optical subclusters \citep{oegerle95}, suggests that the merger is not taking
place in the plane of the sky.  We propose a merger model where the trajectory lies approximately 65$^\circ$ out of the plane of the sky, and at this viewing angle it becomes even more challenging to make quantitative statements about the ICM properties near the cold front.  Nevertheless, we estimate that the pressure change across the front is $\lesssim2.5$, and that the higher density ICM also has the lower temperature.  The estimated merger Mach number of $\sim 1.1$ is likewise consistent with 
other merging systems such as A3667.  We propose a two body merger where A2319B merged from the southeast traveling northwest, with the A2319B galaxies and dark matter passing through the A2319A core roughly 0.4~Gyr ago.

The measured emission-weighted mean temperature of this messy, 
merging cluster is 
$T_{\rm X} = 11.8 \pm 0.6$ keV and the mean abundance 
is $Z = 0.19 \pm 0.03$, using a hydrogen column toward the cluster of
$N_{\rm H} = 8.33\times10^{20}$ cm$^{-2}$. 
The fit values deviate somewhat
from previous studies of A2319. Our higher temperature is likely
due in part to the small field of view of \Chandra\ compared with other
instruments used to study this cluster. Also, we have shown that the emission-weighted mean
 temperature depends sensitively on the choice of energy band, which is at 
least partly explained by the highly nonuniform temperature structure we
have revealed in this cluster.

Our temperature map shows substructure now considered typical in merging 
systems. The cool core of A2319A is readily visible, and the angular separation
of the galaxies of A2319B and an associated cool ICM region indicates a 
separation of the galaxies from the ICM of this subcluster, a transient
phenomenon that gives further evidence of a merger event. There is some 
evidence for a hot bridge of ICM between the two cores, a characteristic associated with shock heating in mergers that has been seen in simulations. 

We examine how this merger affects the bulk properties of A2319. We naively apply the hydrostatic equilibrium assumption to measure a
total mass within \r2500\ of $M_{2500} = (4.2\pm0.5)\times10^{14}~M_\odot$, a factor of 
$1.6\pm0.3$ lower than the mass predicted by a mass--temperature relation
derived from five intermediate-redshift clusters \citep{allen01}. This offset is the strongest indication that the structure of Abell~2319 has been significantly affected by the merger.  Our measurements
for \L2500--temperature, \Micm--temperature, and isophotal size--temperature 
are compared to a sample of 44
clusters observed with \ROSAT\ PSPC \citep{mohr97a,mohr99,mohr00a}. In all 
three cases our measured values for A2319 are within the scatter of the
PSPC-derived scaling relations.  In addition, we note that the $K$-band light in the A2319 galaxy population is consistent with that expected for a cluster of this emission weighted mean temperature \citep{lin03c}.   It is possible that changes
in bulk parameters due to mergers are actually quite large but take place in a correlated way that maintain the low, observed scatter in cluster scaling relations; this has been shown for some properties in numerical simulations \citep{ricker01,evrard02b}. 
However, it will require further studies to determine whether it is possible for mergers to create large, correlated displacements in five cluster parameters (i.e., emission-weighted mean temperature, luminosity, isophotal size, ICM mass, and $K$-band galaxy light) that maintain the low scatter in all four scaling relations.  Another possibility is that despite the X-ray imaging spectroscopy and optical evidence for an ongoing merger in A2319, a merger of this scale is simply not sufficient to grossly perturb the bulk properties of the cluster.

\acknowledgements

We thank Yen-Ting Lin for providing results of a near-infrared analysis of A2319 prior to publication.  We thank an anonymous referee for helpful comments.  This work was supported through the {\it Chandra X-ray Observatory} grant 
G02-3181X and NASA LTSA grant NAG5-11415.   This work made use of a Digitized Sky Survey image.  The Digitized Sky Surveys were produced at the Space Telescope Science Institute under U.S. Government grant NAG W-2166. 

\bibliographystyle{apj}
\bibliography{cosmology,refslock}

\begin{thebibliography}{50}
\expandafter\ifx\csname natexlab\endcsname\relax\def\natexlab#1{#1}\fi

\bibitem[{{Abell}(1958)}]{abell58}
{Abell}, G.~O. 1958, \apjs, 3, 211

\bibitem[{{Allen} {et~al.}(2001){Allen}, {Schmidt}, \& {Fabian}}]{allen01}
{Allen}, S.~W., {Schmidt}, R.~W., \& {Fabian}, A.~C. 2001, \mnras, 328, L37

\bibitem[{{Bialek} {et~al.}(2002){Bialek}, {Evrard}, \& {Mohr}}]{bialek02}
{Bialek}, J.~J., {Evrard}, A.~E., \& {Mohr}, J.~J. 2002, \apjl, 578, L9

\bibitem[{{Buote} \& {Tsai}(1996)}]{buote96}
{Buote}, D.~A., \& {Tsai}, J.~C. 1996, \apj, 458, 27+

\bibitem[{{David} {et~al.}(1993){David}, {Slyz}, {Jones}, {Forman}, {Vrtilek},
  \& {Arnaud}}]{david93}
{David}, L.~P., {Slyz}, A., {Jones}, C., {Forman}, W., {Vrtilek}, S.~D., \&
  {Arnaud}, K.~A. 1993, \apj, 412, 479

\bibitem[{{De Grandi} \& {Molendi}(2001)}]{degrandi01}
{De Grandi}, S., \& {Molendi}, S. 2001, \apj, 551, 153

\bibitem[{{Dickey} \& {Lockman}(1990)}]{dickey90}
{Dickey}, J.~M., \& {Lockman}, F.~J. 1990, \araa, 28, 215

\bibitem[{{Dressler} \& {Shectman}(1988)}]{dressler88}
{Dressler}, A., \& {Shectman}, S.~A. 1988, \aj, 95, 985

\bibitem[{{Evrard} \& {Gioia}(2002)}]{evrard02b}
{Evrard}, A.~E., \& {Gioia}, I.~M. 2002, in ASSL Vol. 272: Merging Processes in
  Galaxy Clusters, 253--304

\bibitem[{{Faber} \& {Dressler}(1977)}]{faber77}
{Faber}, S.~M., \& {Dressler}, A. 1977, \aj, 82, 187

\bibitem[{{Fabricant} {et~al.}(1980){Fabricant}, {Lecar}, \&
  {Gorenstein}}]{fabricant80}
{Fabricant}, D., {Lecar}, M., \& {Gorenstein}, P. 1980, \apj, 241, 552

\bibitem[{{Feretti} {et~al.}(1997){Feretti}, {Giovannini}, \&
  {Bohringer}}]{feretti97}
{Feretti}, L., {Giovannini}, G., \& {Bohringer}, H. 1997, New Astronomy, 2, 501

\bibitem[{{Finoguenov} {et~al.}(2001){Finoguenov}, {Reiprich}, \&
  {B{\"o}hringer}}]{finoguenov01}
{Finoguenov}, A., {Reiprich}, T.~H., \& {B{\"o}hringer}, H. 2001, \aap, 368,
  749

\bibitem[{{Geller} \& {Beers}(1982)}]{geller82}
{Geller}, M.~J., \& {Beers}, T.~C. 1982, \pasp, 94, 421

\bibitem[{{Haiman} {et~al.}(2001){Haiman}, {Mohr}, \& {Holder}}]{haiman01}
{Haiman}, Z.~., {Mohr}, J.~J., \& {Holder}, G.~P. 2001, \apj, 553, 545

\bibitem[{{Hu}(2003)}]{hu03b}
{Hu}, W. 2003, ArXiv Astrophysics e-prints, 1416

\bibitem[{{Irwin} \& {Bregman}(2000)}]{irwin00}
{Irwin}, J.~A., \& {Bregman}, J.~N. 2000, \apj, 538, 543

\bibitem[{{Kempner} {et~al.}(2003){Kempner}, {Sarazin}, \&
  {Markevitch}}]{kempner03}
{Kempner}, J.~C., {Sarazin}, C.~L., \& {Markevitch}, M. 2003, \apj, 593, 291

\bibitem[{{Kempner} {et~al.}(2002){Kempner}, {Sarazin}, \&
  {Ricker}}]{kempner02}
{Kempner}, J.~C., {Sarazin}, C.~L., \& {Ricker}, P.~M. 2002, \apj, 579, 236

\bibitem[{{Landau} \& {Lifshitz}(1987)}]{landau87}
{Landau}, L.~D., \& {Lifshitz}, E.~M. 1987, {Fluid Mechanics} (2nd ed.; Oxford:
  Butterworth-Heinemann)

\bibitem[{{Lin} {et~al.}(2003){Lin}, {Mohr}, \& {Stanford}}]{lin03c}
{Lin}, Y., {Mohr}, J.~J., \& {Stanford}, S.~A. 2003, \apj, submitted

\bibitem[{{Lockman}(2003)}]{lockman03}
{Lockman}, F.~J. 2003, in Conference Proceedings, in preparation

\bibitem[{{Majumdar} \& {Mohr}(2003{\natexlab{a}})}]{majumdar03a}
{Majumdar}, S., \& {Mohr}, J.~J. 2003{\natexlab{a}}, \apj, 585, 603

\bibitem[{{Majumdar} \& {Mohr}(2003{\natexlab{b}})}]{majumdar03b}
---. 2003{\natexlab{b}}, \apj, submitted (astroph/0305341)

\bibitem[{{Markevitch}(1996)}]{markevitch96}
{Markevitch}, M. 1996, \apjl, 465, L1

\bibitem[{{Markevitch} {et~al.}(1998){Markevitch}, {Forman}, {Sarazin}, \&
  {Vikhlinin}}]{markevitch98}
{Markevitch}, M., {Forman}, W.~R., {Sarazin}, C.~L., \& {Vikhlinin}, A. 1998,
  \apj, 503, 77+

\bibitem[{{Markevitch} {et~al.}(2002){Markevitch}, {Gonzalez}, {David},
  {Vikhlinin}, {Murray}, {Forman}, {Jones}, \& {Tucker}}]{markevitch02}
{Markevitch}, M., {Gonzalez}, A.~H., {David}, L., {Vikhlinin}, A., {Murray},
  S., {Forman}, W., {Jones}, C., \& {Tucker}, W. 2002, \apjl, 567, L27

\bibitem[{{Markevitch} {et~al.}(2003){Markevitch}, {Mazzotta}, {Vikhlinin},
  {Burke}, {Butt}, {David}, {Donnelly}, {Forman}, {Harris}, {Kim}, {Virani}, \&
  {Vrtilek}}]{markevitch03}
{Markevitch}, M., {Mazzotta}, P., {Vikhlinin}, A., {Burke}, D., {Butt}, Y.,
  {David}, L., {Donnelly}, H., {Forman}, W.~R., {Harris}, D., {Kim}, D.-W.,
  {Virani}, S., \& {Vrtilek}, J. 2003, \apjl, 586, L19

\bibitem[{{Markevitch} {et~al.}(2000){Markevitch}, {Ponman}, {Nulsen}, {Bautz},
  {Burke}, {David}, {Davis}, {Donnelly}, {Forman}, {Jones}, {Kaastra},
  {Kellogg}, {Kim}, {Kolodziejczak}, {Mazzotta}, {Pagliaro}, {Patel}, {Van
  Speybroeck}, {Vikhlinin}, {Vrtilek}, {Wise}, \& {Zhao}}]{markevitch00}
{Markevitch}, M., {Ponman}, T.~J., {Nulsen}, P. E.~J., {Bautz}, M.~W., {Burke},
  D.~J., {David}, L.~P., {Davis}, D., {Donnelly}, R.~H., {Forman}, W.~R.,
  {Jones}, C., {Kaastra}, J., {Kellogg}, E., {Kim}, D.~., {Kolodziejczak}, J.,
  {Mazzotta}, P., {Pagliaro}, A., {Patel}, S., {Van Speybroeck}, L.,
  {Vikhlinin}, A., {Vrtilek}, J., {Wise}, M., \& {Zhao}, P. 2000, \apj, 541,
  542

\bibitem[{{Markevitch} \& {Vikhlinin}(2001)}]{markevitch01b}
{Markevitch}, M., \& {Vikhlinin}, A. 2001, \apj, 563, 95

\bibitem[{{Maughan} {et~al.}(2003){Maughan}, {Jones}, {Ebeling}, {Perlman},
  {Rosati}, {Frye}, \& {Mullis}}]{maughan03}
{Maughan}, B.~J., {Jones}, L.~R., {Ebeling}, H., {Perlman}, E., {Rosati}, P.,
  {Frye}, C., \& {Mullis}, C.~R. 2003, \apj, 587, 589

\bibitem[{{Mohr} \& {Evrard}(1997)}]{mohr97a}
{Mohr}, J.~J., \& {Evrard}, A.~E. 1997, \apj, 491, 38

\bibitem[{{Mohr} {et~al.}(1995){Mohr}, {Evrard}, {Fabricant}, \&
  {Geller}}]{mohr95}
{Mohr}, J.~J., {Evrard}, A.~E., {Fabricant}, D.~G., \& {Geller}, M.~J. 1995,
  \apj, 447, 8+

\bibitem[{{Mohr} {et~al.}(1993){Mohr}, {Fabricant}, \& {Geller}}]{mohr93}
{Mohr}, J.~J., {Fabricant}, D.~G., \& {Geller}, M.~J. 1993, \apj, 413, 492

\bibitem[{{Mohr} {et~al.}(1999){Mohr}, {Mathiesen}, \& {Evrard}}]{mohr99}
{Mohr}, J.~J., {Mathiesen}, B., \& {Evrard}, A.~E. 1999, \apj, 517, 627

\bibitem[{{Mohr} {et~al.}(2000){Mohr}, {Reese}, {Ellingson}, {Lewis}, \&
  {Evrard}}]{mohr00a}
{Mohr}, J.~J., {Reese}, E.~D., {Ellingson}, E., {Lewis}, A.~D., \& {Evrard},
  A.~E. 2000, \apj, 544, 109

\bibitem[{{Molendi} {et~al.}(1999){Molendi}, {de Grandi}, {Fusco-Femiano},
  {Colafrancesco}, {Fiore}, {Nesci}, \& {Tamburelli}}]{molendi99}
{Molendi}, S., {de Grandi}, S., {Fusco-Femiano}, R., {Colafrancesco}, S.,
  {Fiore}, F., {Nesci}, R., \& {Tamburelli}, F. 1999, \apjl, 525, L73

\bibitem[{{Nagai} \& {Kravtsov}(2003)}]{nagai03}
{Nagai}, D., \& {Kravtsov}, A.~V. 2003, \apj, 587, 514

\bibitem[{{Oegerle} {et~al.}(1995){Oegerle}, {Hill}, \& {Fitchett}}]{oegerle95}
{Oegerle}, W.~R., {Hill}, J.~M., \& {Fitchett}, M.~J. 1995, \aj, 110, 32

\bibitem[{{Onuora} {et~al.}(2003){Onuora}, {Kay}, \& {Thomas}}]{onuora03}
{Onuora}, L.~I., {Kay}, S.~T., \& {Thomas}, P.~A. 2003, \mnras, 341, 1246

\bibitem[{{Pearce} {et~al.}(1994){Pearce}, {Thomas}, \& {Couchman}}]{pearce94}
{Pearce}, F.~R., {Thomas}, P.~A., \& {Couchman}, H.~M.~P. 1994, \mnras, 268,
  953

\bibitem[{{Randall} {et~al.}(2002){Randall}, {Sarazin}, \&
  {Ricker}}]{randall02}
{Randall}, S.~W., {Sarazin}, C.~L., \& {Ricker}, P.~M. 2002, \apj, 577, 579

\bibitem[{{Ricker} \& {Sarazin}(2001)}]{ricker01}
{Ricker}, P.~M., \& {Sarazin}, C.~L. 2001, \apj, 561, 621

\bibitem[{{Roettiger} {et~al.}(1997){Roettiger}, {Loken}, \&
  {Burns}}]{roettiger97}
{Roettiger}, K., {Loken}, C., \& {Burns}, J.~O. 1997, \apjs, 109, 307

\bibitem[{{Struble} \& {Rood}(1987)}]{struble87}
{Struble}, M.~F., \& {Rood}, H.~J. 1987, \apjs, 63, 543

\bibitem[{{Sun} {et~al.}(2002){Sun}, {Murray}, {Markevitch}, \&
  {Vikhlinin}}]{sun02}
{Sun}, M., {Murray}, S.~S., {Markevitch}, M., \& {Vikhlinin}, A. 2002, \apj,
  565, 867

\bibitem[{{Townsley} {et~al.}(2000){Townsley}, {Broos}, {Garmire}, \&
  {Nousek}}]{townsley00}
{Townsley}, L.~K., {Broos}, P.~S., {Garmire}, G.~P., \& {Nousek}, J.~A. 2000,
  \apjl, 534, L139

\bibitem[{{Vikhlinin} {et~al.}(2001){Vikhlinin}, {Markevitch}, \&
  {Murray}}]{vikhlinin01}
{Vikhlinin}, A., {Markevitch}, M., \& {Murray}, S.~S. 2001, \apj, 551, 160

\bibitem[{{Vikhlinin} {et~al.}(2002){Vikhlinin}, {VanSpeybroeck}, {Markevitch},
  {Forman}, \& {Grego}}]{vikhlinin02}
{Vikhlinin}, A., {VanSpeybroeck}, L., {Markevitch}, M., {Forman}, W.~R., \&
  {Grego}, L. 2002, \apjl, 578, L107

\bibitem[{{Zabludoff} \& {Zaritsky}(1995)}]{zabludoff95}
{Zabludoff}, A.~I., \& {Zaritsky}, D. 1995, \apjl, 447, L21+

\end{thebibliography}

\end{document}